\documentclass[prb,showpacs,twocolumn,superscriptaddress]{revtex4}
\usepackage{amsmath,graphicx,latexsym}

\begin{document}

\title{\emph{Ab-initio} theory of metal-insulator interfaces in a finite 
       electric field}

\author{Massimiliano Stengel}
\author{Nicola A. Spaldin}

\affiliation{Materials Department, University of California, Santa Barbara,
             CA 93106-5050, USA}

\date{\today}

\begin{abstract}
We present a novel technique for calculating
the dielectric response of metal/insulator heterostructures.
This scheme allows, for the first time, the fully first-principles 
calculation of the microscopic properties of thin-film capacitors 
at finite bias potential. 
The method can be readily applied to pure insulators, 
where it provides an interesting alternative
to conventional finite-field techniques based on the
Berry-phase formalism.
We demonstrate 
the effectiveness of our method by performing comprehensive numerical 
tests on a model Ag/MgO/Ag heterostructure.
\end{abstract}


\maketitle


\section{Introduction}

The dielectric response of a metal-insulator interface to an applied
electric field is a subject of great interest nowadays, as the ongoing 
miniaturization of electronic devices is reaching the atomic
scale.
In this regime, the properties of thin oxide films 
(used e.g. in nonvolatile ferroelectric memories~\cite{Ahn/Rabe/Triscone:2004} 
and as gate oxides in MOSFET transistors~\cite{Eisenbeiser_et_al:2000}) 
start to deviate from those predicted by macroscopic models, and cannot 
be disentangled from the metallic or semiconducting 
contact~\cite{Dawber/Rabe/Scott:2005}.
Traditionally, deviations from bulk dielectric behavior in thin film
geometries have been described using phenomenological models, which have
provided valuable insights into the \emph{qualitative} consequences
of low dimensionality and interfacing to a metallic electrode. However,
such models do not have predictive \emph{quantitative} capabilities for
the many physical factors, such as interface
chemistry~\cite{Giustino/Umari/Pasquarello:2003}, finite screening length
in the metal~\cite{Junquera/Ghosez:2003}, or the presence of structural
and point defects related to the growth process, that contribute to the
observed properties.
In order to separate these effects and to help improve over
the existing models, the electronic and ionic response of a 
metal-insulator heterostructure to an external bias potential must be
understood at the quantum mechanical level.

First-principles calculations, especially within the framework of density 
functional theory, have proven to be a powerful tool in the fundamental
understanding of metal-ceramic interfaces~\cite{Finnis:1996}.
On the other hand,
{\em ab-initio} studies of interfacial dielectric properties are still
in their infancy, and only \emph{insulating} heterostructures
have been successfully simulated~\cite{Giustino/Umari/Pasquarello:2003} in  
a finite external field.
Indeed, when one of the two materials is a metal (as required for
the simulation of a realistic thin-film capacitor), the presence of
partially filled bands at the Fermi level prevents the
straightforward application of the Berry-phase 
technique~\cite{King-Smith/Vanderbilt:1993}, 
on which most finite electric field calculations 
have been based so far~\cite{Souza/Iniguez/Vanderbilt:2002,
Umari/Pasquarello:2002}.
In order to gain further insight into the rich domain of phenomena occuring
at metal-dielectric junctions there is the clear need for a methodology
that allows one to work around this obstacle.

In this work we demonstrate how this problem can be solved
by using techniques and ideas borrowed from
Wannier-function theory, which is an appealing alternative to 
the traditional Berry-phase formalism~\cite{Resta:2002}.
In particular, we show that the insulating nature of 
a capacitor under an applied bias potential is reflected at
the microscopic level in the confinement of the metal-induced gap states, 
whose propagation in the insulator is forbidden; this property allows for 
a rigorous definition of the macroscopic polarization in real space.
The coupling to an external field is then introduced through the usual
definition of the electric enthalpy, in close analogy to purely insulating 
systems~~\cite{Souza/Iniguez/Vanderbilt:2002,
Umari/Pasquarello:2002}.

We demonstrate the effectiveness of our technique using a model MgO/Ag(100) 
heterostructure; the simple properties of this well studied metal-ceramic 
interface make it an ideal test case for our method.
In addition to providing a technical validation to our approach, 
our results indicate that MgO films retain an ideal classical
behavior in the ultrathin limit, corroborating 
the picture that emerged from previous investigations~\cite{Schintke:2001}.

This work is organized as follows. In Section II we define the polarization,
the electric enthalpy and the Hamiltonian for our new finite-field approach,
and show how it can be used to calculate the capacitance and local permittivity
profiles of a general metal/insulator heterostructure.
In Section III we relate our scheme to other possible approaches.
In Section IV we present the numerical test on Ag/MgO in full detail, 
with particular focus on the new properties of the system
that are made accessible by our scheme (dynamical charges at the interface,
electronic response, etc.).
In Section V we briefly discuss the applicability of our method to range of
physical problems, and present the summary and conclusions.

\section{Method}

To place our developments in context, we begin by reviewing the discrete 
Berry-phase formalism introduced by King-Smith and 
Vanderbilt~\cite{King-Smith/Vanderbilt:1993} (KSV)
for a general insulating crystal; we then adapt the method we developed in 
Ref.~\onlinecite{Stengel/Spaldin:2005} to obtain a real-space expression for
the polarization. 
%
%
Next (Sec.~\ref{sechet}) we show how this procedure can be readily 
extended to periodic metal/insulator heterostructures. 
Finally, in Sec.~\ref{seccou} we introduce the coupling to an 
external field in the total energy and the Hamiltonian, which completes 
the derivation of our finite-field method.
A discussion of the first-principles capacitance per unit area is provided
in Sec.~\ref{seccap}, and links our method to the theory of local 
permittivity~\cite{Giustino/Pasquarello:2005},
which was originally developed for purely insulating heterostructures.

\subsection{Insulating systems}
\label{secins}
%
We start with a periodic insulator described by three real space lattice
vectors $\mathbf{R}_i$; for convenience, we impose that 
$\mathbf{R}_1=(a,0,0)$ is perpendicular to $\mathbf{R}_{2,3}$ 
(the latter two lie therefore on the $yz$ plane). 
The Brillouin zone for such a Bravais lattice is a prism in
reciprocal space defined by the reciprocal space vectors 
$\mathbf{G}_1=(2\pi / a,0,0)$ and $\mathbf{G}_{2,3}$.
We choose a discrete $k$-point sampling of the type $\mathbf{k}=j\mathbf{b} +
\mathbf{k}_\perp$, where the vector 
$\mathbf{b_/parallel}= (2\pi/L,0,0) = \mathbf{G}_1 / N_\parallel$ spans a
regular one-dimensional mesh of dimension $N_\parallel = L/a$ and
$\mathbf{k}_\perp$ belongs to a set of $N_\perp$ special points in the 
perpendicular plane; 
the total number of independent $k$-points is then $N_\perp N_\parallel$.
The $x$ component of macroscopic polarization $\langle P \rangle$ 
is commonly evaluated on this reciprocal space mesh by first defining 
a set of matrices:
\begin{equation}
M^{\mathbf{k}}_{mn} = \langle u_{m\mathbf{k}} |
                               u_{n\mathbf{k}+\mathbf{b}_\parallel} \rangle,
\end{equation}
where it is understood that 
$|u_{n \mathbf{k}+\mathbf{G}_\parallel} \rangle=
 e^{-i \mathbf{G}_\parallel  \mathbf{r}} |u_{n\mathbf{k}} \rangle$.
In the KSV approach~\cite{King-Smith/Vanderbilt:1993} $\langle P \rangle$ 
is then calculated as a discrete Berry phase for each individual 
``stripe'' of $k$-points:
\begin{equation}
\langle P \rangle_{Berry}(\mathbf{k}_\perp) = 
-\frac{2e}{\Omega} \frac{a}{2\pi}
\sum_j
\mathrm{Im} \ln \det M^{\mathbf{k}_\perp + j\mathbf{b}};
\label{eqberry}
\end{equation}
the resulting values are finally averaged over the set of special points
$\mathbf{k}_\perp$ with corresponding weights $w_{{k}_\perp}$:
\begin{equation}
 \langle P \rangle_{Berry} = \sum_{{k}_\perp} w_{{k}_\perp}  
 \langle P \rangle_{Berry}(\mathbf{k}_\perp).
\end{equation}

As an alternative, within single-particle theories, 
the macroscopic polarization can be equivalently related to the 
centers of the Wannier functions~\cite{King-Smith/Vanderbilt:1993}. 
In Ref.~\onlinecite{Stengel/Spaldin:2005} we have introduced a two-step 
procedure to obtain accurate values of the polarization by using 
a real-space technique based on Wannier functions:
\begin{itemize}
\item First, the maximally localized Wannier 
functions~\cite{Marzari/Vanderbilt:1997} are determined iteratively
starting from a set of ground-state Bloch orbitals;
\item Then, the charge distribution of the Wannier function is constructed
in real space, and the center defined by a saw-tooth function.
\end{itemize}
In the following, we modify this strategy to operate in one dimension 
instead of three, for several reasons: i) this way we can achieve an 
even more transparent analogy between the Wannier-function real-space method 
and the KSV Berry-phase formula; ii) an external field couples to one 
component of the polarization only; iii) it can be generalized in a
very natural way to a system which is at the same time a two-dimensional 
conductor and a one-dimensional insulator (see Section~\ref{sechet}).

When the first step above is restricted to one dimension only, the 
resulting ``hermaphrodite'' Wannier 
functions~\cite{Sgiarovello/Peressi/Resta:2001,
Giustino/Pasquarello:2005} are maximally localized along the $x$ 
direction and Bloch-like in the orthogonal $yz$ plane and are much easier to 
obtain compared to their three-dimensional counterparts.
In fact, the unitary transformation that yields such a set of localized orbitals
is constructed by means of the parallel-transport technique in a one-shot 
singular value decomposition, without the need for the iterative minimization
of the spread functional~\cite{Marzari/Vanderbilt:1997}.
The parallel transport technique 
consists~\cite{Marzari/Vanderbilt:1997} of enforcing the particular
representation of the orbitals 
in which $M^\mathbf{k}_{mn}=K^\mathbf{k}_{mn} \lambda^{\mathbf{k}_\perp}_n$,
where $\lambda^{\mathbf{k}_\perp}_n$ are complex numbers of the form
$\lambda^{\mathbf{k}_\perp}_n = e^{i\phi_n}$ and $K^\mathbf{k}$ is Hermitian.
A definition of the Wannier centers can be written in terms
of the phases $\phi_n$:
\begin{equation}
\bar{x}_n^{\mathbf{k}_\perp} = 
\frac{a N_\parallel}{2\pi} \phi_n.
\label{trigcen}
\end{equation}
Interestingly, this definition of the centers of the parallel-transported 
hermaphrodite Wannier functions bears an exact relationship with
the KSV \emph{discrete} Berry-phase polarization: 
\begin{equation}
\langle P \rangle_{Berry}(\mathbf{k}_\perp) = 
  \frac{2e}{\Omega} \sum_n \bar{x}_n^{\mathbf{k}_\perp}.
\end{equation}

%
%
Once the approximate values, $\bar{x}_n^{\mathbf{k}_\perp}$, of the Wannier 
centers are available, the ``refinement'' in real space by means of saw-tooth
functions (second step above) can be carried out in close analogy to 
Ref.~\onlinecite{Stengel/Spaldin:2005}.
%
%
First, the charge density associated with the hermaphrodite Wannier function 
$|w_{n\mathbf{k}_\perp} \rangle$ is explicitly constructed in a supercell 
which is a multiple of the primitive one (of total length $L$ along $x$) and  
integrated over the $yz$ planes to yield a one-dimensional function
of periodicity $L$:
\begin{equation}
\rho_n^{\mathbf{k}_\perp}(x) = \int dy dz 
|w_n^{\mathbf{k}_\perp}(\mathbf{r})|^2,
\end{equation}
Then, the refined Wannier centers $x_n$ are defined as:
\begin{equation}
x_n^{\mathbf{k}_\perp} = 
\int_{\bar{x}_n^{\mathbf{k}_\perp}-L/2}^{\bar{x}_n^{\mathbf{k}_\perp}+L/2} 
x \rho_n^{\mathbf{k}_\perp}(x) dx. 
\label{refinedcen}
\end{equation}
Finally, the expression for the polarization $\langle P \rangle$ 
follows:
\begin{equation}
\langle P \rangle (\mathbf{k}_\perp)= \frac{2e}{\Omega} 
\sum_n x_n^{\mathbf{k}_\perp}. 
\label{eqpins}
\end{equation}

The advantages of using the real-space refinement step described above are
twofold:
\begin{itemize}
\item The real-space centers $x_n^{\mathbf{k}_\perp}$ are identical to the
Berry-phase centers $\bar{x}_n^{\mathbf{k}_\perp}$ in the 
limit of $N_\parallel \rightarrow \infty$, but for a discrete mesh they
differ; in particular, $x_n^{\mathbf{k}_\perp}$ converge \emph{exponentially}
to the asymptotic limit, while $\bar{x}_n^{\mathbf{k}_\perp}$ have a much
slower, polynomial convergence 
(as $\mathcal{O}(N_\parallel^{-2})$)~\cite{Stengel/Spaldin:2005}. Therefore
in practice a relatively coarse $k$-point mesh is sufficient for the
calculation of $P$, with a significantly reduced computational effort.
\item The real-space expression for the polarization allows for a
straightforward extension to metal/insulator heterostructures
and the introduction of the coupling to a finite external field,
which is the main goal of this work and will be treated in depth
in the next subsection.
\end{itemize}

\subsection{Metal/insulator heterostructures}

\label{sechet}

\begin{figure}
\includegraphics[width=0.5\textwidth]{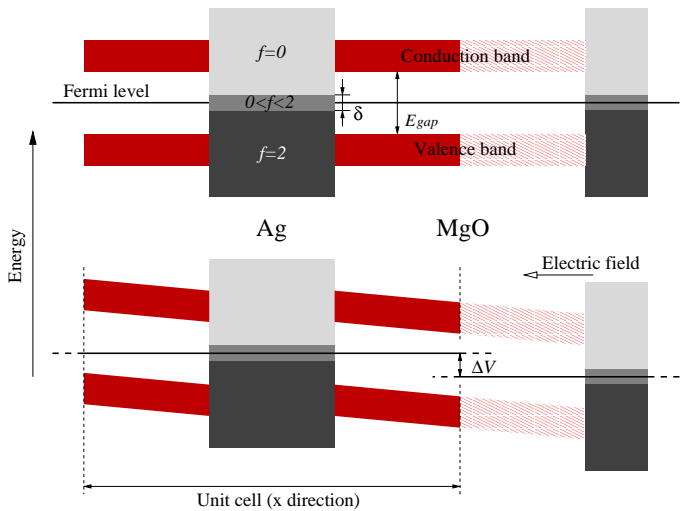}
\caption{Schematic representation of the 
projected band structure of a metal-insulator heterostructure in the
absence of
(top) and with (bottom) electric field. An effective bias potential 
$\Delta V$  between one slab and the repeated images is obtained.}
\label{fig1}
\end{figure}

We remind the reader that, in a metal/insulator heterostructure,
the usual reciprocal-space expression for $\langle P \rangle$ 
based on the Berry phase is incompatible with the
fractional orbital occupancies which stem from the metallicity
of the electrodes; however,
we will show hereafter that the real-space expression of the polarization 
described above can be adapted to provide a rigorous description
also in this case.
We start from the same assumptions and definitions as in the insulating
case, except that we set $N_\parallel=1$, i.e. only the two-dimensional
special-points average on $\mathbf{k}_\perp$ is performed in the 
Brillouin zone sampling. 
This is by no means a limitation, since for a physically meaningful study
of a polarized capacitor a necessary condition is that there be no coupling
between the electrodes across the dielectric film, i.e. the corresponding 
one-particle bands must be virtually dispersionless along the $x$ direction.

Incidentally, we note that ``hermaphrodite'' WFs, which are the basic ingredient
of the method, are ideally suited to representing the inherently 
delocalized nature of electrons in the two-dimensionally conducting plates, 
while providing a fast spatial decay (which is crucial to an 
accurate definition of $\langle P \rangle$~\cite{Stengel/Spaldin:2005}) 
along the field direction ($x$).
Unfortunately, the ``localization'' transformation, upon which the real-space
formulation is rooted, cannot be performed straightforwardly here because
of the fractional occupancies at the Fermi level: in the finite-temperature
functional the total energy is no longer invariant upon a unitary 
transformation of the occupied states~\cite{Marzari/Vanderbilt/Payne:1997}.
To work around this problem and achieve a localized representation
along $x$, we divide the problem into three regimes, which (as above) 
are identified independently for every $\mathbf{k}_\perp$ point of the 
surface Brillouin zone (a schematic representation is shown in 
Fig.~\ref{fig1}):
\begin{itemize}
\item
The conduction states with zero occupancy
are discarded from the computation, since they do not contribute 
to the ground state energy or charge density.
\item
The submanifold of fully occupied valence states (with energy eigenvalue
$\epsilon_n<E_F-\delta$ in Fig.~\ref{fig1}) is localized separately 
by means of the parallel-transport 
algorithm~\cite{Marzari/Vanderbilt:1997,umesh}; this operation leaves 
the total energy unaffected in complete analogy to the purely
insulating case.
\item
The partially occupied states in the region of the Fermi level (shown as 
``$0<f<2$'' in Figure~\ref{fig1}) lie in the energy gap of the 
dielectric; as a result they are quantum mechanically confined
in the metallic slab without taking any further action, since they cannot
propagate through the insulator.
\end{itemize}

The centers, $x_n^{\mathbf{k}_\perp}$, of the resulting set of localized orbitals are then
calculated, analogously to the insulating case, by using Eq.~\ref{refinedcen}
(for the metal-induced gap states we set $\bar{x}_n=0$ and assume by
convention that the metallic slab is approximately centered at $x=0$).
%
The definition of the polarization is identical to the
insulating case, except that we must take 
the orbital occupation numbers $f_n^{\mathbf{k}_\perp}$ explicitly 
into account:
\begin{equation}
\langle P \rangle (\mathbf{k}_\perp) = \frac{e}{\Omega} 
\sum_n f_n^{\mathbf{k}_\perp} 
x_n^{\mathbf{k}_\perp}.
\label{eqpx}
\end{equation}
The multi-valuedness in the definition of $x_n^{\mathbf{k}_\perp}$ (modulo a 
lattice vector) is removed by imposing $x_n^{\mathbf{k}_\perp} \in [-L/2,L/2]$, 
with the origin $x=0$ corresponding to the center of the metallic slab and $L$ the 
cell length~\footnote{
  According to the modern theory of polarization, only differences of $\langle P \rangle$
  between two states of the system connected by a continuous transformation are
  observable quantities, while the absolute value of $\langle P \rangle$ is only defined
  modulo a \emph{quantum of polarization}~\cite{King-Smith/Vanderbilt:1993}.
  The condition $x_n^{\mathbf{k}_\perp} \in [-L/2,L/2]$ (with the
  metallic slab approximately centered in the origin), is important to  
  ensure that all fractionally occupied state belong to the same supercell;
  in this case it is easy to show that $\langle P \rangle$ is well defined and
  has the same properties as in the insulating case.}.  
Since the band energies change their value during structural relaxation,
the whole procedure has to be repeated at the beginning of every iteration.
This means that the number of states assigned to each energy window is not 
fixed during the simulation, but is free to vary, following the evolution 
of orbital eigenvalues. 

We note that the exact boundaries of the energy windows, in particular 
the energy separating the
orbitals that are considered metal-induced gap states and those that are
localized by means of the parallel-transport procedure, is to some extent
arbitrary; this choice, however, has no influence on the result
(provided that in the lower window all states are fully occupied and
that the middle window does not overlap with the conduction or valence band 
of the insulator). The method is therefore well defined.

\subsection{Electric enthalpy and Hamiltonian}

\label{seccou}

The coupling of a periodic system to a finite external field $\mathcal{E}$ is 
described, in both the metallic and insulating cases, 
by the electric 
enthalpy~\cite{Umari/Pasquarello:2002,Souza/Iniguez/Vanderbilt:2002}:
\begin{equation}
E^\mathcal{E} = E_{KS} - \Omega \mathcal{E}.\langle P \rangle,
\end{equation}
where $E_{KS}$ is the Kohn-Sham total energy, $\Omega$ is the 
volume of the supercell and $\langle P \rangle$
is the macroscopic polarization along the field axis; 
in a parallel-plate capacitor this is always 
perpendicular to the electrode plane.

We write the contribution $\Delta \hat{H}^\mathcal{E}$ to 
the Hamiltonian due to the electric field operator as a 
non-local, state-dependent potential $F_n(x)$ applied to each WF, 
$w_n$, individually (we drop the superscript $\mathbf{k_\perp}$
in the following discussion): 
\begin{equation}
\Delta \hat{H}^\mathcal{E} = \mathcal{E} 
\sum_n |\hat{F}_n w_n \rangle \langle w_n|.
\label{operator}
\end{equation}
The $F_n(\mathbf{r})$ are periodic sawtooth functions~\footnote{
  The $F_n(\mathbf{r})$ are constructed in reciprocal space in order to
  avoid Fourier aliasing errors and preserve analytical gradients.} 
  of the form
$F_n(\mathbf{r})=x$, with $x\in [x_n-L/2,x_n+L/2]$, i.e. $F_n(0,y,z)=0$ for every
$n$, and is discontinuous in a point located far from the center $x_n$, 
where the corresponding orbital has vanishing importance.
We note that the average value of $F_n(x)$ over the cell is 
irrelevant for a purely insulating system, but is crucial when a
metallic electrode is present. 
It can be easily verified that our choice respects the correct limit in 
the case of an isolated metallic slab in vacuum, where the field operator 
becomes {\em local}. 
This also means that, when a center $x_n$ crosses the point $x=L/2$ 
(the center of the insulating slab), the corresponding polarizing 
potential $\mathcal{E} F_n(x)$ undergoes a discontinuous, 
rigid jump of magnitude $\Delta V = \pm \mathcal{E}L$.  
For a sufficiently thick insulating layer, and as long as $\Delta V$ is 
small enough not to bring valence or conduction states into the 
partially occupied region close to the Fermi level, this jump leaves 
the system unaffected apart from a trivial shift of 
the eigenvalues corresponding to fully-occupied Bloch states
(the discontinuity can also be effectively smoothened by smearing 
the $F_n$ with Gaussian functions; a practical procedure to do this 
is explained in the Appendix).
Interestingly, the large-field Zener instability reported for purely insulating
bulk materials~\cite{Souza/Iniguez/Vanderbilt:2002,
Umari/Pasquarello:2002} translates here to a {\em Schottky-tunneling}
instability that has to be avoided by careful analysis of the band 
alignment.

$\Delta \hat{H}^\mathcal{E}$ has in general a small anti-Hermitian part
due to the fact that the Wannier functions are never exactly zero at the
cell boundary but retain a small (exponentially vanishing) tail:
\begin{displaymath}
\langle w_i |\Delta \hat{H}^\mathcal{E}| w_j \rangle -
(\langle w_j |\Delta \hat{H}^\mathcal{E}| w_i \rangle)^* =
\langle w_i | (F_j - F_i) | w_j \rangle.
\end{displaymath}
The influence of this anti-Hermitian component is to induce a ``unitary
torque'' within the manifold of the occupied
orbitals, which is undesirable since, 
when working in the ensemble density functional theory
framework (or even more clearly in the case of insulating systems), 
the total energy (and the electric enthalpy) are both invariant with
respect to such a unitary transformation; this is therefore a spurious 
contribution that we remove by introducing a correction term of the form:
\begin{equation}
\Delta \hat{H}'|w_i \rangle = 
  \frac{\mathcal{E}}{2} 
  \sum_j |w_j \rangle \langle w_j | ( F_j - F_i ) | w_i \rangle.
\label{eqcorrec}
\end{equation}
This term goes exponentially to zero with increasing insulator thickness,
and vanishes if each Wannier-like state is strictly
localized in $x \in [x_n-L/4,x_n+L/4]$.
With this correction, we could always achieve optimal convergence to the 
electronic ground state.

%

We want to stress here that, at the origin of the non-zero
anti-Hermitian term (Eq.~\ref{eqcorrec}), there's an important
\emph{physical} feature, i.e. the tunneling current across a thin 
insulating film is never exactly zero (although it is exponentially
decreasing as a function of film thickness).
In order to study the metastable (long-lived) 
state of the polarized capacitor within a ground-state theory, we 
are obliged to make an approximation and discard the tunneling current 
by truncating the tails of the evanescent conduction states induced by
the metallic electrodes.
%
We expect this approximation to be very good in cases where the tunneling
current is so negligibly small that it does not contribute appreciably to the 
physics of the system.
This condition is indeed satisfied in the regime where our method
is exact, i.e. when the tails of the metal-induced gap states are 
confined to a region much smaller than the thickness of the insulating film,
and therefore direct tunneling is suppressed.

%

\subsection{Local dielectric response and capacitance}

\label{seccap}

The theory of local permittivity that was originally 
developed for purely insulating interfaces~\cite{Giustino/Pasquarello:2005} 
is directly applicable to metal/insulator heterostructures, and can be 
readily generalized to introduce the first-principles capacitance 
density per unit area, $\mathcal{C}$.
We start from the usual textbook definition of the capacitance density,
\begin{equation}
\mathcal{C} = \frac{\sigma}{\Delta V},
\label{classcap}
\end{equation}
with $\Delta V=\mathcal{E}L$ the potential drop between the electrodes, and
$\sigma$ the surface density of the \emph{free} charge stored in the device.
From classical electrostatics, the free charge is equal to the 
value of the induced polarization deep inside the electrode, 
where the applied field is completely screened.
In our geometry this occurs in the center ($x=0$) of the metallic slab, and
Eq.~\ref{classcap}
can be rewritten as
\begin{equation}
\mathcal{C} = \frac{\Delta \bar{P}(x=0)}{\mathcal{E}L},
\end{equation}
where $\Delta \bar{P}(x)$ is the planar- and 
macroscopically-averaged~\cite{Baldereschi:1988} induced
polarization.
Finally, $\Delta \bar{P}(0)$ can be extracted from the relationship, derived in
Ref.~\onlinecite{Giustino/Pasquarello:2005}, between the local permittivity,
$\bar{\epsilon}(x)$, and the induced polarization, and
the corresponding globally averaged quantities $\langle \epsilon \rangle$ and
$\langle \Delta P \rangle$:
\begin{equation}
\Big( 1 - \frac{1}{\bar{\epsilon}(x)} \Big) =
\Big( 1 - \frac{1}{\langle \epsilon \rangle} \Big) \frac{\Delta
\bar{P}(x)}{\langle \Delta P \rangle} .
\label{locperm}
\end{equation}
Since deep in the metal $\bar{\epsilon}(x)$ diverges [i.e.
$1/\bar{\epsilon}(0)=0$],
Eq.~\ref{locperm} reduces simply to
$\Delta \bar{P}(x=0) = \langle \Delta P \rangle + \mathcal{E}/(4\pi)$,
and Eq.~\ref{classcap} reduces to
\begin{equation}
\mathcal{C} = \frac{\langle \epsilon \rangle}{4\pi L},
\label{capa}
\end{equation}
where $\langle \epsilon \rangle=4\pi \langle \Delta P \rangle / \mathcal{E} + 1$
is the
overall dielectric constant of the supercell, and $\langle \Delta P \rangle$ is
the
polarization induced by the field $\mathcal{E}$. We use
equation~\ref{capa} to obtain the capacitance values 
in Section~\ref{sectest}.

In order to appreciate the influence of the interface on the properties of the device,
it is often interesting to compare the \emph{ab-initio} value of the
capacitance (Eq.~\ref{capa}) to the well-known
classical formula, which is valid in the macroscopic 
regime and depends only on the thickness $t$ 
and on the bulk permittivity of the dielectric film $\epsilon_{bulk}$:
\begin{equation}
\mathcal{C}_{classical}
  =  \frac{\epsilon_{bulk}}{4\pi t}.
\label{cclass}  
\end{equation}   
In a size regime comparable to the electronic wavelength the 
``thickness'' of the insulating slab ceases to be a rigorous concept,
since the transition from one bulk material to the other 
is not sharp, and their individual properties are strongly modified
in close vicinity of the junction. 
In most cases we can define a ``nominal'' thickness as $t=mt_0$, 
where $m$ is the number of layers of the crystalline film and $t_0$
is the equilibrium interlayer spacing of the dielectric film far from the
interface. 
Together with this definition of the thickness, we will use Eq.~\ref{cclass} 
in Section~\ref{sectest} to define a nominal capacitance $\mathcal{C}_0$.
%

\section{Alternative approaches}
\begin{figure}
\includegraphics[width=0.45\textwidth]{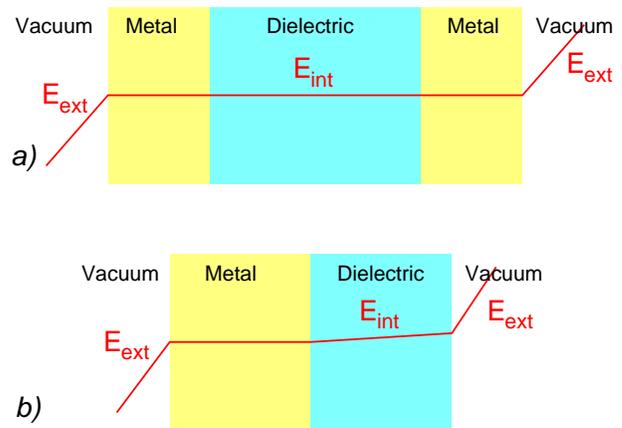}
\caption{Schematic representation of the self-consistent electrostatic
potential (red curve) in unsupported metal/insulator/metal (a) and 
metal/insulator (b) slabs when an external field $\mathcal{E}_{ext}$ is 
switched on. In the geometry (a) by classical electrostatics the internal
macroscopic field acting on the dielectric $\mathcal{E}_{int}$ is zero;
in case (b) it corresponds to $\mathcal{E}_{ext}/\epsilon_{bulk}$, where
$\epsilon_{bulk}$ is the bulk dielectric constant of the insulator.}
\label{figalt}
\end{figure}
\subsection{Metal/insulator heterostructures}

In principle, to address the problem of the dielectric transition at a 
metal/insulator interface, several alternatives to our method can be explored;
we will briefly review in this section the most obvious solutions.

We have already mentioned that the Berry phase formalism is difficult to 
apply because of the partially occupied states at the Fermi level.
Conventional linear-response~\cite{Baroni:2001} techniques seem also problematic, 
because the response to a uniform external field is again incompatible 
with the overall metallic character of the system. 
A promising alternative is provided by quantum transport
techniques, either within the nonequilibrium Green's function formalism 
or within time-dipendent density functional theory~\cite{Burke/Car/Gebauer:2005}; our 
approach should yield comparable results in the limit of small 
tunneling current, although probably at a substantially lower computational 
cost.

A different approach consists in introducing a layer of vacuum in the 
heterostructure, in order to allow for the straightforward application of a 
local saw-tooth potential; since a capacitor is a finite object this seems also
a natural choice for the geometry (Fig.~\ref{figalt}a). 
%
%
%
%
%
%
%
As the two electrodes lie in the same supercell, however, there is only one 
Fermi level that defines the chemical potential of an electron in both
metallic slabs; in other words, an external perturbation
can not induce a potential difference between the capacitor plates.
Instead, the system behaves as a short circuit,
analogously to a thick, homogeneous metallic slab, irrespective of
the dielectric material sandwiched in between. In this case, the external field 
is completely
screened by the accumulation/depletion of charge at the free surfaces of
the conductor.
%
%

A possibility to work around this problem is to simulate just
``half'' of the capacitor, i.e. a metal/insulator slab isolated in vacuum
(Fig.~\ref{figalt}b).
Again, an external saw-tooth function can be used both to apply a uniform external field
and to define the polarization $\langle P \rangle$; from
the self-consistent electronic ground state in
this perturbed potential we can extract charge density differences
and local polarization profiles, that can then be used to 
estimate the capacitance and interfacial properties.
There are, however, some limitations to this approach:
\begin{itemize}
\item The termination of the insulating lattice must not be metallic
or we recover case 1 above;
this rules out the simulation of most polar 
interfaces [e.g. (100)-oriented 5-1 or 3-3 perovskites]. 
Furthermore, even for ``insulating'' crystal surfaces,
sometimes surface states due to dangling bonds substantially reduce the
electronic gap, thus limiting the range of electric field
values that can be used in practice.
\item By classical electrostatics, the electric field in the vacuum is 
larger than the electric field in the insulator by a factor of $\epsilon$.
This means that, for high-permittivity dielectrics,
simulations are restricted to extremely small values of the internal field, 
and therefore the non-linear regime is not practically accessible by means 
of this strategy.
Even worse, when the insulator is a ferroelectric, it has already been shown
that the presence of a free surface makes it difficult to impose the correct
electrical boundary conditions inside the film~\cite{Meyer_Vanderbilt}.
\end{itemize}

The advantage of our method is that its limitations are exclusively 
dictated by fundamental physics (Schottky or direct tunneling), not by 
artifacts that must be introduced \emph{ad hoc} in an unconventional 
simulation cell.
We note that the band alignment at the metal/dielectric junction plays
a crucial role in determining the properties of the interface; particular 
care must be used to assess the reliability of the exchange and correlation 
functional in reproducing this feature, at least at a qualitative level.

\subsection{Pure insulators}

A real-space approach for applying a finite external field to a periodic system 
was first proposed by Nunes and Vanderbilt~\cite{Nunes/Vanderbilt:1994} 
and applied to a tight-binding Hamiltonian, which was based on the 
linear-scaling scheme of Mauri, Galli and Car~\cite{Mauri/Galli/Car:1993}.
The first application of this approach to a self-consistent 
density-functional calculation was presented by Fern\'andez, {Dal~Corso} and 
A.~Baldereschi~\cite{Fernandez/DalCorso/Baldereschi:1998}, and validated 
by the study of the dielectric properties of bulk Si and GaAs.
%
The great advantage provided by the linear scaling functional 
(Ref.~\onlinecite{Mauri/Galli/Car:1993}) 
is that it leads naturally to a description of the electronic structure 
in terms of Wannier functions which are strictly confined in space; 
this means that the real-space position operator is formally well-defined.
The price to pay, however, is that one has to deal with a non-conventional 
functional, where the orthonormality constraint is not exactly satisfied 
and that might pose convergence problems during the search 
for the electronic ground state; we are not aware of practical follow-ups 
of these techniques.

The vast majority of the finite-field calculations reported so far 
were performed within the scheme of Souza, \'{I}\~{n}iguez
and Vanderbilt~\cite{Souza/Iniguez/Vanderbilt:2002} or Umari and
Pasquarello~\cite{Umari/Pasquarello:2002}, both based on the Berry-phase 
definition of polarization.
In the case of pure insulators, our scheme can be considered as an 
interesting practical alternative to the latter two approaches,
since it can yield substantially better accuracy for the same choice of
computational parameters.

\section{Application and numerical tests}
\label{sectest}
%
%

\begin{figure}
\includegraphics[width=0.45\textwidth]{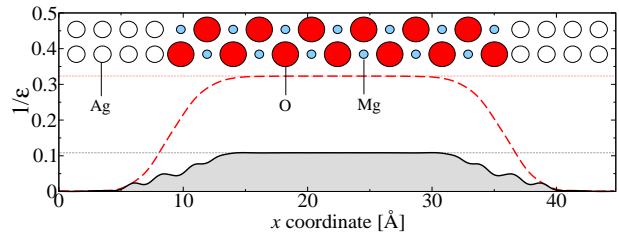}
\caption{High-frequency (red dashed curve) and static (black shaded curve)
inverse permittivity profiles for the Ag/MgO heterostructure. Bulk values
for the inverse permittivity are indicated as thin dotted lines. 
The approximate
positions of the atomic planes are sketched in the upper part of the figure.}
\label{fig2}
\end{figure}

We demonstrate the effectiveness of our scheme by 
presenting results for a model [Ag(100)]$_n$/[MgO(100)]$_m$ 
heterostructure, with the thicknesses of the Ag and MgO slabs set to 
$n=9$ and $m=13$ atomic layers respectively. 
We choose to fix the in-plane lattice constant to our calculated equilibrium
value of bulk MgO ($a_0$=7.86 a.u.), and apply a 3\% uniform in-plane 
expansion to the Ag lattice in order to obtain a pseudomorphic structure.
The O atoms are placed on top of the interface Ag atoms, which is the
most favorable alignment; the interface Mg atoms then sit above a
four-fold hollow site of the Ag(100) surface.

\subsection{Computational parameters and zero-field equilibrium structure}

  Our calculations were performed within the local density 
  approximation~\cite{Perdew/Wang:1992} and the PAW method~\cite{Bloechl:1994}
  as implemented in our ``in-house'' code,
  with a plane wave cutoff of 40 Ry, a Gaussian smearing of 0.2 eV
  for the orbital occupancies and 10 special $k$-points in the
  surface Brillouin zone.
  For all electronic and ionic relaxations we used an extension of the
  Car and Parrinello~\cite{Car/Parrinello:1985} method to metallic 
  systems~\cite{joost,lagrange}.
The out-of-plane atomic positions and cell parameter were relaxed in zero field
until the maximum force was lower than 0.05 meV/\AA, and the spacing $t_0$ 
between the central MgO layers was within 0.2\% of the bulk equilibrium value.

\subsection{High-frequency response}

We then imposed an external field of 51.4 V/$\mu$m and relaxed the electrons 
to the ground state while keeping the ions fixed. The resulting high-frequency 
permittivity profile is shown as a thick dashed curve in Fig.~\ref{fig2}. 
On the metal side of the interface the inverse permittivity drops quickly 
to zero, as expected, while it is very close to our calculated bulk MgO value
after only three atomic layers on the insulator side 
(the bulk calculations were performed on a tetragonal 8-layer 
non-primitive supercell with the same in-plane lattice parameter and
$k$-point sampling as the interface system; 
the out-of-plane spacing was set to $t_0$). 
The calculated value in the middle of the MgO slab is
$\bar{\epsilon}^\infty(L/2)$=3.095, which is within 0.03\% of our calculated 
bulk value, $\epsilon^\infty_{bulk}$=3.094. 
This level of accuracy 
reflects the favorable convergence of the real-space expression of
the polarization, which was demonstrated in Ref.~\onlinecite{Stengel/Spaldin:2005}.
As a comparison, in a previous 
calculation of insulating heterostructures based on the Berry-phase approach,
errors introduced solely by the finite field formalism were estimated 
to be of the order of 10\%~\cite{Giustino/Pasquarello:2005}.
The overall dielectric constant of the heterogeneous supercell was
$\langle \epsilon^\infty \rangle$ = 5.203.

\subsection{Dynamical charges}

\begin{figure}
\includegraphics[width=0.45\textwidth]{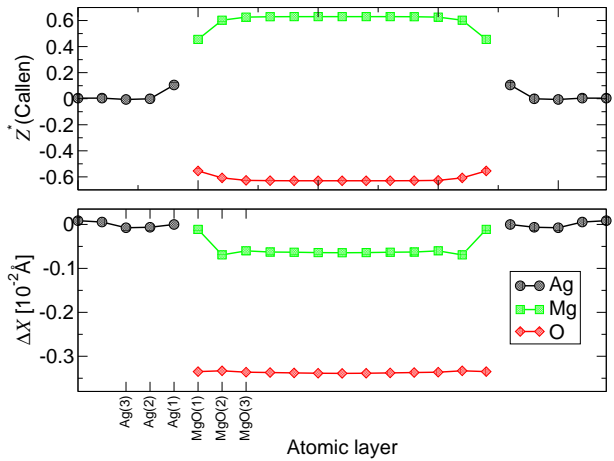}
\caption{(i) $x$ (longitudinal) component of the Callen dynamical charges 
in the Ag/MgO heterostructure in units of $e$. 
(ii) Atomic relaxations induced by a field of $10^{-4}$ a.u.}
\label{figz}
\end{figure}

\begin{table}
\begin{tabular}{|c|c|}
\hline
Layer & $Z^*$(Callen) \\ 
\hline
Ag(4)  &  0.004 \\
Ag(3) & -0.006 \\
Ag(2) & -0.001 \\
Ag(1) & 0.106 \\
MgO(1) & 0.455, -0.554 \\
MgO(2) & 0.603, -0.608 \\
MgO(3) & 0.626, -0.627 \\
MgO(4) & 0.629, -0.629 \\
\hline
Bulk MgO & 0.630, -0.630 \\
\hline
\end{tabular}
\caption{Callen dynamical charges at the MgO/Ag(100) interface; in MgO 
the first values refers to Mg, the second to O.}
\label{dynch}
\end{table}

The ionic forces $f_I$ extracted from the finite-field ground state 
calculation provide the Born effective charges $Z^*_I=f_I/\mathcal{E}$ 
of all atoms at once~\cite{Umari/Pasquarello:2002}; their values,
however, are not a local property of the interface as they depend on the
overall geometry and composition of the supercell.
It is physically more insightful, instead, to look at the Callen 
(or longitudinal) dynamical charges, which are obtained by dividing the
Born charges by the average high-frequency dielectric constant of the 
heterostructure $\langle \epsilon^\infty \rangle$, and are indeed 
local~\cite{sio2_ir}.
The Callen charges for the present system are represented in the upper 
panel of Fig.~\ref{figz}; their values for the atoms closest to the 
interface are reported in Tab.~\ref{dynch}.
It is apparent that,
already three or four layers away from the interface, the charges 
associated to the Mg and O ions have already converged to the bulk value;
their absolute value is slightly reduced 
in the first two MgO layers, while
at the same time the interfacial Ag atom has a non-negligible positive 
charge.
This behavior is consistent with the penetration of the metallic states 
in the insulator, which partially screens the ionic response of the ions
closest to the interface; at the same time, the external field has a finite
penetration length in the metallic slab, which affects the interface Ag layer.
%
%
%

\subsection{Static response}

The ionic contribution to the static 
polarization can be either evaluated in the linear (zero field) approximation
with standard linear-response techniques~\cite{Baroni:2001}, or 
non-perturbatively, by performing a direct relaxation in a finite external
field~\cite{Umari/Pasquarello:2002}; we tested both approaches here.
First we computed the symmetry-restricted force matrix of the system 
by the finite-difference method in zero field using small displacements 
of 0.0004 \AA{} along $x$. 
The variation of the \emph{local}
polarization upon atomic displacements, $\delta \bar{P}(x,\{R_I\}) / \delta R_I$, was
also simultaneously extracted for each degree of freedom and used,
together with the force matrix and the Born effective charges, to construct 
the static permittivity profile, $\bar{\epsilon}^0(x)$, which is shown as a 
thick solid curve in Fig.~\ref{fig2}. 
The behavior of the static and high frequency permittivities are similar,
the only relevant difference being their magnitudes in the middle of
the insulating slab, with $\bar{\epsilon}^0(L/2)$ converging to a value
of 9.227 (our calculated $\epsilon^0_{bulk} = 9.229$).
In the lower panel of Fig.~\ref{figz} we show the field-induced atomic
displacements (the center of the metallic slab has been assumed as a 
reference). 
As expected,
the metallic slab remains practically unperturbed, while interestingly,
with this choice of origin, both Mg and O
sublattices relax opposite to the field direction; the relaxation of the Mg 
ions is much less important than the relaxation of the oxygens.

Next, we performed a direct relaxation of the atomic positions
in an external bias $\Delta V=1.15$ V, 
which corresponds roughly to a self-consistent field of 400 V/$\mu$m 
throughout the dielectric; the resulting average permittivity was within 
0.05\% of the zero field linear response value, indicating as expected 
the absence of anharmonic effects in this field regime within numerical 
accuracy.

\subsection{Capacitance density}

Having assessed the reliability of our technique, and obtained accurate
profiles, we next extracted the static and high-frequency values of the 
capacitance, obtaining $\mathcal{C}^0=31.2$ fF/$\mu$m$^2$ and 
$\mathcal{C}^\infty=10.3$ fF/$\mu$m$^2$ respectively.
These values are respectively 3.1\% and 1.2\% higher than the classical 
estimates based only on the knowledge of the bulk permittivity and of the 
thickness of the film:
\begin{equation}
\mathcal{C}^{(0,\infty)}_{classical}
  =  \frac{\epsilon^{(0,\infty)}_{bulk}}{4\pi mt_0}
\end{equation}   
The close similarity between $\mathcal{C}$ and $\mathcal{C}_{classical}$ 
indicates that the influence of the metal on the dielectric properties 
of the MgO film is very small. 
Previous investigations of ultrathin MgO films deposited on Ag have already 
evidenced a surprisingly fast recovery of the bulk oxide local electronic 
structure as a function of film thickness~\cite{Schintke:2001}; 
here we see that this result also holds 
for the \emph{dielectric} properties of MgO films, which show a
nearly ideal classical behaviour down to the nanometer scale. 

\subsection{Molecular dynamics}

\begin{figure}
\includegraphics[width=0.45\textwidth]{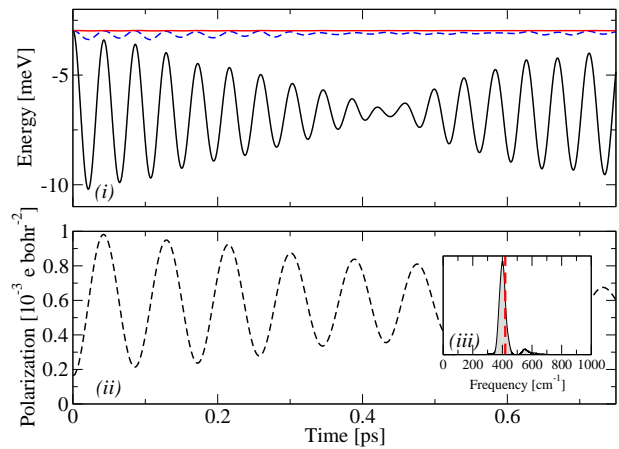}
\caption{(i) Energy conservation during molecular dynamics in an external 
bias of 1.15 V. Represented are $E^\mathcal{E}$ (black solid), 
$E^\mathcal{E}+K_{ion}$ (blue dashed) and $E^\mathcal{E}+K_{ion}+K_{fake}$
(red solid).
The zero of the energy scale is set at the ground state total energy 
in zero field. (ii) Polarization, $\langle P \rangle$, 
as a function of time. (iii) Frequency-resolved dipolar activity of the
system (black shaded curve); zone-center transverse optical phonon 
frequency of bulk MgO is also shown (red dashed line).}
\label{fig3}
\end{figure}

Since the formalism is not exactly variational, it is important 
to assess its suitability for calculating
finite-temperature effects by monitoring energy conservation 
during a microcanonical molecular dynamics run~\cite{joost}. 
We start from the centrosymmetric zero-field equilibrium structure and 
let the atoms evolve along the $x$ axis under an external bias
of $\Delta V=1.15$ V.
The evolution of the potential energy $E^\mathcal{E}$, of the ``physical'' 
constant of motion $E^\mathcal{E}+K_{ion}$ and of the approximate 
``mathematical'' constant of motion $E^\mathcal{E}+K_{ion}+K_{fake}$ 
($K_{ion}$ is the ionic kinetic energy, while $K_{fake}$ is the fictitious 
electronic kinetic energy~\cite{Car/Parrinello:1985}) are shown in
the upper panel of Fig.~\ref{fig3}. 
The quality of the energy conservation is excellent; the curve has
no appreciable drift, even when enlarging the vertical scale by 
three orders of magnitude.
In the lower panel we show the evolution of $\langle P \rangle$, which 
undergoes quasiperiodic oscillations of approximate frequency 390 cm$^{-1}$. 
To link the molecular dynamics run to the lattice-dynamical properties of
the system extracted from the force matrix, in the inset we show the phonon 
spectrum broadened by a Gaussian of width 20 cm$^{-1}$ and weighted by the 
dipolar activity of the mode, $[Z^*(\omega)]^2$, where $Z^*(\omega)$ is the 
mode effective charge~\cite{Antons:2005}.
The dominant peak, centered at 400 cm$^{-1}$, closely corresponds to the 
main oscillation period of the polarization $\langle P \rangle$ during the 
dynamics; the frequency of the calculated zone-center transverse
optical mode of bulk MgO (420 cm$^{-1}$) is indicated as a thick dashed line.

\section{Conclusions}

In summary, we have developed a first-principles method for calculating
the response of hybrid metal/insulator structures to an electric field, and
have demonstrated its applicability by calculating the local permittivity
and capacitance of a model thin-film capacitor. 
The method is also appropriate for insulating materials at finite bias, 
where it provides greater accuracy than conventional techniques, in both 
the linear and anharmonic regimes with respect to field and temperature; 
this is crucial in the simulation of ferroelectric devices, where 
anharmonic effects are an essential physical ingredient for a realistic 
description~\cite{Antons:2005}.
The generality of the technique is evidenced by the high quality of
overall energy conservation in molecular dynamics, which suggests 
fascinating applications to phenomena occuring at liquid/electrode
interfaces, e.g. the recently observed freezing of 
interfacial water under electric fields~\cite{Choi:2005}.

This work was supported by the National Science Foundation's Division of
Materials Research through the Information Technology Research program,
grant number DMR-0312407, and made use of MRL Central Facilities
supported by the National Science Foundation under award No. DMR-0080034

\appendix

\section{Numerical implementation}

In this Appendix we will describe the technical details of the implementation
of our method within the projector-augmented wave~\cite{Bloechl:1994} (PAW) 
formalism. The same prescriptions apply to the closely related ultrasoft 
pseudopotential technique~\cite{VanderbiltUS},
while the case of norm-conserving pseudopotentials is easily recovered by 
removing the augmentation terms in the following expressions.

\subsection{Notation}

The basic ingredients of the PAW method are a set of 
all-electron (AE) and pseudo (PS) partial waves, respectively 
$|\phi_{\tau,i}\rangle$ and $|\tilde{\phi}_{\tau,i}\rangle$, 
and the corresponding
projector functions $|\tilde{p}_{\tau,i}\rangle$.
The index $i$ here embeds three indices: two angular momentum variables 
$l$ and $m$, and a third one, $n$, that accounts for the fact that several partial
waves per angular momentum can be used; the dependence on the atomic 
species/site is included in $\tau$.
It is useful to define two kind of integrals based on these quantities, 
namely the overlap matrix elements:
\begin{equation}
Q_{\tau,ij} = \langle  \phi_{\tau,i} | \phi_{\tau,j}\rangle -
\langle \tilde{\phi}_{\tau,i} | \tilde{\phi}_{\tau,j}\rangle,
\end{equation}
and the localized dipole moments of the augmented charges:
\begin{equation}
d_{\tau,ij} = \langle  \phi_{\tau,i} |x| \phi_{\tau,j}\rangle -
\langle \tilde{\phi}_{\tau,i} |x| \tilde{\phi}_{\tau,j}\rangle.
\end{equation}
Most of these integrals are zero because of the angular momentum 
selection rules. 
The projectors are used to construct the local density matrices:
\begin{equation}
W_{\tau,ij} = \sum_{n\mathbf{k}} f_{n\mathbf{k}} 
  \langle \psi_{n\mathbf{k}}|\tilde{p}_i \rangle
  \langle \tilde{p}_j | \psi_{n\mathbf{k}} \rangle
\end{equation}

\subsection{Polarization}

Next, following Ref.~\onlinecite{notesdhv} we first separate the total polarization
into three parts:
\begin{equation}
P=P_{ion}+P^{BP}+P^{EV},
\end{equation}
respectively the ionic contribution, the Berry phase (BP) contribution 
and the expectation value (EV) term. The first is trivial, and is given by
the dipole moment of the bare pseudo-atomic charges:
\begin{equation}
P_{ion} = \frac{e}{\Omega} \sum_\tau Q_\tau X_\tau.
\end{equation} 
The third (EV) term is also easy to compute and is nothing but the total 
dipole moment of the augmented charges:
\begin{equation}
P^{EV} = -\frac{e}{\Omega} \sum_{\tau,ij} d_{\tau,ij} W_{\tau,ij}.
\end{equation}

The second (BP) term has the same form as Eq.~\ref{eqberry}:
\begin{equation}
\langle P \rangle_{Berry}^{BP}(\mathbf{k}_\perp) =
-\frac{2e}{\Omega} \frac{a}{2\pi}
\sum_j
\mathrm{Im} \ln \det M^{\mathbf{k}_\perp + j\mathbf{b}};
\end{equation}
however, the $M$ matrix here must include the appropriate augmentation terms:
$$
M^{\mathbf{k}}_{mn} = \langle u_{m\mathbf{k}} |
                               u_{n\mathbf{k}+\mathbf{b}_\parallel} \rangle +
$$
\begin{equation}			       
\sum Q_{\tau,ij} \langle u_{m\mathbf{k}} | \tilde{p}^{(\mathbf{k})}_i \rangle
\langle \tilde{p}^{(\mathbf{k+b_\parallel})}_j| u_{n\mathbf{k}+\mathbf{b}_\parallel} \rangle
\end{equation}

The parallel transport construction explained in Sec.~\ref{secins} then 
provides the unitary rotations that localize the Wannier functions along the 
field axis, which lead easily to the one-dimensional PS Wannier densities 
$\tilde{\rho}_n(x)$.
Within PAW, however, there is one supplementary step that must be taken 
when explicitly constructing the localized orbitals in real space, 
i.e. one must reconstruct the localized augmented charges associated with
the Wannier orbital; in particular we will need in the following the
\emph{total} augmented charge per site associated to the $n$-th
Wannier function:
\begin{equation}
q_{\tau,n} = \sum_{ij} Q_{\tau,ij} \langle w_n | \tilde{p}_i \rangle
\langle \tilde{p}_j |w_n \rangle
\end{equation}

\subsection{Saw-tooth functions}

Finally, to complete our definition of the polarization we need to construct
the saw-tooth functions $F_n(x)$ in real space; in the following we present a
recipe to do so while ensuring a continuous evolution of the Hamiltonian 
operator. 
Given a Wannier center $w_n$ of approximate center $\bar{x}_n$ (we always set 
$\bar{x}_n=0$ for the metal-induced gap states with fractional orbital occupancies),
defined in a one-dimensional cell of length $L$, we construct first the 
periodic function:
\begin{equation}
H_n(x) = 1 - A\sum_{l=-\infty}^\infty e^{-[x-\bar{x}_n+(l+1/2)L]^2/\sigma^2} 
\end{equation}
where $A=L/(\sqrt{\pi} \sigma)$ 
is such that the average value of $H$ is zero and $\sigma$ is 
about 0.5-1 bohr. Then we define our saw-tooth function as the integral
of $H_n$:
\begin{equation}
F_n(x) = \int_0^x H_n(y) dy
\end{equation}
We note that this definition ensures that $F_n(0)=0$ for all $n$; 
the finite value of $\sigma$ effectively avoids any discontinuity in the
evolution of the Hamiltonian while facilitating the plane-wave representation.
Then the ``refined'' center $x_n$ is defined as:
\begin{equation}
x_n = \int_0^L F_n(x) \tilde{\rho}_n(x) dx + 
\sum_\tau F_n(X_\tau) q_{\tau,n} + \bar{x}_n - F_n(\bar{x}_n), 
\end{equation}
where $q_{\tau,n}$ is the augmented charge at the atomic site $\tau$ 
associated with the $n$-th Wannier function, and $X_\tau$ is the atomic 
position. Note that the charges $q_{\tau,n}$ and the PS densities satisfy
the usual generalized normalization condition:
\begin{equation}
\int_0^L \tilde{\rho}_n(x) dx + \sum_\tau q_{\tau,n} = 1
\end{equation}

The $F_n(x)$ functions are then used in Eq.~\ref{operator} to build the
electric field contribution to the Hamiltonian.

\bibliography{Max}

\end{document}